\documentclass[reprint,amsmath,superscriptaddress,unsortedaddress,amssymb,aps,prb]{revtex4-1}

\usepackage{graphicx}

\usepackage{psfrag}

\newcommand{\bs}{\boldsymbol}
\newcommand{\eps}{\varepsilon}

\newcommand{\be}{\begin{equation}}
\newcommand{\ee}{\end{equation}}
\newcommand{\bea}{\begin{eqnarray}}
\newcommand{\eea}{\end{eqnarray}}

\def\bea{\begin{eqnarray}}
\def\eea{\end{eqnarray}}
\def\be{\begin{equation}}
\def\ee{\end{equation}}

\begin{document}

\title{Crossover in the two-impurity Kondo model induced by direct charge tunneling}

\author{Justin Malecki }
\affiliation{University of British Columbia, Vancouver, BC, Canada V6T1Z1}

\author{Eran Sela}
\affiliation{Institute for Theoretical Physics, University of Cologne, 50937 Cologne, Germany}

\author{Ian Affleck}
\affiliation{University of British Columbia, Vancouver, BC, Canada V6T1Z1}

\begin{abstract}
Quantum critical behavior in the two-impurity Kondo model requires the distinct separation of two scales, $T_K \gg T^*$, where $T_K$ is the Kondo temperature and $T^*$ is the scale at which the system renormalizes away from the quantum critical point to a stable Fermi liquid fixed point.  We provide a derivation of $T^*$ based on the renormalization group to lowest order.  This result is confirmed by a numerical renormalization group (NRG) analysis which supplements the analytic derivation with additional quantitative precision.  The form of the low-energy Fermi liquid fixed point is derived and subsequently confirmed by the NRG.  We discuss implications for series double quantum dot systems.
\end{abstract}

\maketitle

\section{Introduction}
Competing orders in correlated electron systems (\textit{e.g.}\ heavy fermion compounds or high temperature superconductors) lead to new exotic quantum critical points (QCPs). Interestingly, QCPs also occur in much simpler scenarios where correlations are driven fundamentally by a local quantum impurity coupled to a free Fermi sea. One of the most fascinating classes of impurity models showing quantum criticality are the multichannel Kondo models~\cite{NozieresBlandin}. At criticality, the electrons are not described by a Fermi liquid theory, as has been demonstrated in quantum dot experiments~\cite{Potok07}.  However, one element of these critical systems that requires further understanding is the nature of  the crossover away from the QCP. This crossover will generically occur due to arbitrarily small symmetry breaking perturbations (such as the presence of a magnetic field in the multichannel Kondo effect) at temperatures small compared to the crossover scale associated with those perturbations.

Therefore, to obtain closer contact with experiment, an accurate estimate of the crossover energy scale is required. In this paper, we describe such a calculation for a model closely related to multi-channel Kondo models, namely,  the 2-impurity Kondo model which has possible realizations in double quantum dot devices.\cite{Georges,Zarand2006} Related experiments on quantum dots\cite{Jeong} or using a scanning tunneling microscope
with one impurity on the tip and one on the sample~\cite{Wahl} did not observe the physics of the QCP so far.  Indeed, theoretical predictions of the influence of the QCP on the nonlinear conductance~\cite{Sela2009} and shot noise~\cite{Sela09Noise}  require the crossover energy to be small relative to the Kondo temperature.  One of the central purposes of this paper is to study the effect of the simplest symmetry breaking perturbation in the two impurity Kondo model, namely direct charge tunneling, via a numerical calculation of the crossover scale. We shall briefly discuss the applicability to double quantum dot setups.

We consider two leads labelled by $L$ and $R$, each coupled to one of two spin-1/2 ``impurities'' labelled by $\bs{S}_L$ and $\bs{S}_R$ respectively.  The two spins are coupled via a Heisenberg exchange interaction $K$.  We also consider a direct charge tunneling term $V_{LR}$ between the two leads.
The system is described by the model Hamiltonian
\bea
\nonumber H & = & \int_{-\infty}^\infty dx \sum_{j = L,R} \psi_{j\mu}^\dag(x) i \partial_x \psi_{j \mu}(x) \nonumber \\&&+ J \left(\bs{s}_{LL} \cdot \bs{S}_L + \bs{s}_{RR} \cdot \bs{S}_R \right) \\ \nonumber
& & + K \bs{S}_L \cdot \bs{S}_R + V_{LR} \left( \psi_{L \mu}^\dag (0) \psi_{R \mu}(0) + \psi_{R \mu}^\dag(0) \psi_{L \mu}(0) \right).
\label{2IKH}\eea
The operator $\psi_{j\mu}(x)$ creates a chiral electron of spin $\mu$ at position $x$ in the $j^\mathrm{th}$ lead (these chiral operators are defined on the entire real line via the standard ``unfolding'' transformation).  We have defined the spin operators $\bs{s}_{jj} := \psi_{j\mu}^\dag(0) \frac{\bs{\sigma}_{\mu \nu}}{2} \psi_{j \nu}(0)$ ($\bs{\sigma}$ are the Pauli matrices) which couple to each impurity spin.

For the special case of $V_{LR}=0$, it is known~\cite{Jones} that a quantum critical point separating two distinct Fermi liquid phases exists for a critical value of $K = K_c \sim T_K$.  $T_K$ is the Kondo temperature which we take to be
\be
\label{Tk}
T_K = D \sqrt{\nu J} e^{- \frac{1}{\nu J}}
\ee
where $2D$ is the bandwidth or ultra-violet cutoff and $\nu =1/(2\pi)$ is the density of states in the leads at the Fermi energy (recall that, in our units, the Fermi velocity is $v_F = 1$).  Throughout this paper, we set $\hbar =k_B = 1$.  For $K=K_c$, the effective low-energy model cannot be described as a Fermi liquid.  We use the following nomenclature to describe the three phases of the model at $V_{LR}=0$:
\bea
K < K_c & & \quad \textrm{Kondo screened phase (KSP)} \nonumber  \\
K = K_c & & \quad \textrm{quantum critical point (QCP)} \nonumber \\
K > K_c & & \quad \textrm{local singlet phase (LSP)} \nonumber
\eea

For $K$ close to $K_c$ and $V_{LR}$ close to $0$ (the meaning of ``close'' will be defined shortly), the effective description of the system will depend on the temperature $T$.  For $T \gg T_K$, the spins are only weakly coupled to the leads.  For $T^* \ll T \ll T_K$, the system will be described by the QCP while, for $T \ll T^*$, the system will be described by a Fermi liquid that is continuously related to the KSP or the LSP via the parameter $V_{LR}$ as will be described in detail in \S~\ref{sec:FL}.

It has been proposed~\cite{Sela2009} that the temperature scale $T^*$ can be estimated as
\be
\label{tstar}
T^* = a \frac{(K - K_c)^2}{T_K} + b T_K (\nu V_{LR})^2
\ee
where we have inserted two dimensionless numbers $a$ and $b$ that are expected to be of order unity.   The value of $T^*$ determines how close $K$ must be to $K_c$ and how close $V_{LR}$ must be to $0$ in order to observe the QCP: the values must be such that $T^* \ll T_K$.

To understand where this estimate for $T^*$ comes from, let us assume that $K$ is tuned close to $K_c$ and that $\nu V_{LR}$ is
very small. We want to estimate the crossover energy scale at which the system renormalizes away from
the critical point, due to both $\delta K \equiv K-K_c$ and also $\nu V_{LR}$. To make a rough estimate, we
use the lowest order weak coupling renormalization group (RG) at energy scales above $T_K$ and the RG at the QCP below $T_K$.   At weak coupling, $\delta K$ has dimension 1 and $\nu V_{LR}$ is
dimensionless. Thus, at scale $T_K$,
\bea
\delta K(T_K)&\approx&  (D/T_K)\delta K\nonumber \\
\nu V_{LR}(T_K)&\approx &\nu V_{LR}.
\eea
Here, the quantities on the right are the bare ones at scale $D$ and the quantities on the left
are the renormalized ones at scale $T_K$. At energy scales $E<T_K$, both
$\delta K$ and $\nu V_{LR}$ have dimension 1/2 so
\bea
\delta K(E)&\approx & (D/T_K)(T_K/E)^{1/2}\delta K\nonumber \\
\nu V_{LR}(E)&\approx & (T_K/E)^{1/2}\nu V_{LR}.
\eea

We estimate the crossover energy scale $T_{\delta K}$ associated with the coupling $\delta K$ (not to be confused with the Kondo temperature $T_K$)
by setting $\delta K(T_{\delta K})/D\approx 1$.   Similarly, we estimate the crossover energy scale $T_{LR}$ associated with the potential
scattering by setting $\nu V_{LR}(T_{LR})\approx 1$.  Thus
\bea
T_{\delta K}&\approx& a (\delta K)^2/T_K \nonumber \\
\label{TdKTLR}
T_{LR}&\approx& b(\nu V_{LR})^2T_K
\eea
where the dimensionless factors $a$ and $b$ are used to account for quantitative details missed by this simplified analysis. The total crossover scale is then given by $T^* = T_{\delta K} + T_{LR}$.
We might expect further modifications if we took into account
higher order terms in the RG equations at weak coupling and in the vicinity of
the QCP.  Hopefully this just leads to corrections which are logarithmic
in the dimensionless parameters.

The estimate of eq.~(\ref{tstar}) agrees with similar analytic arguments~\cite{Zarand2006, Georges1999} (although it is slightly smaller than that found in the latter reference).  In \S~\ref{sec:nrgtstar} we verify this estimate by a transparent and systematic NRG study with the aim of estimating this crossover scale.  We find satisfactory agreement with eq.~(\ref{tstar}) though with the magnitude of the coefficients $a$ and $b$ differing from unity.  We also find evidence that the coefficient $b$ may have some residual dependence on the Kondo coupling $J$ that is not explained by the simple scaling analysis that lead to eq.~(\ref{tstar}).

In \S~\ref{sec:FL}, we derive the Fermi liquid theory for the stable low-energy fixed point which is parametrized by the value of $V_{LR}$.  The resulting theory is described in terms of a boundary condition at the origin which can be related to a scattering phase shift.  The formula for this phase shift is predicted analytically and supported by comparison with the phase shift derived from the NRG.
In \S~\ref{sec:disc} we
discuss the two impurity Anderson model version of this problem, showing that  the simplified Kondo model of Eq. (\ref{2IKH}) is
{\it not} the correct low energy approximation to the Anderson model.

\section{Estimating $T^*$ using the NRG}
\label{sec:nrgtstar}
The numerical renormalization group is a powerful, non-perturbative algorithm for studying quantum impurity systems.  It was originally developed to study single-impurity model in [\onlinecite{wilson1975, krishnamurthy1980a, krishnamurthy1980b}] where the technique is exhaustively described.  It has since been applied to numerous other impurity problems~\cite{bulla2008} including the two-impurity problem that we study here~\cite{Jones}.  We refer the reader to these references for details on the implementation of the NRG and only review those elements necessary for calculating $T^*$.

The key idea of the NRG is to approximate the original Hamiltonian describing the leads by that of two semi-infinite tight-binding ``Wilson chains''
\bea
\nonumber\frac{H}{D} & = & \frac{1}{2} \left(1 + \Lambda^{-1} \right) \sum_{n=0}^\infty \sum_{p=e,o} \Lambda^{-\frac{n}{2}} \xi_n \left( f_{np\mu}^\dag f_{n+1,p \mu} + \textrm{h.c.} \right) \\
\nonumber & & + \nu J \sum_{p = e,o} f_{0 p \mu}^\dag \frac{\bs{\sigma}_{\mu \nu}}{2} f_{0 p \nu} \cdot \left( \bs{S}_L + \bs{S}_R \right) \\
\nonumber & & + \nu J \left( f_{0 e \mu}^\dag \frac{\bs{\sigma}_{\mu \nu}}{2} f_{0 o \nu} + \textrm{h.c.} \right) \cdot \left( \bs{S}_L - \bs{S}_R \right) \\
& & + \frac{K}{D} \bs{S}_L \cdot \bs{S}_R + 2 \nu V_{LR} \left( f_{0e\mu}^\dag f_{0 e \mu} - f_{0 o \mu}^\dag f_{0 o \mu} \right).
\eea
Each of the $f_{np\mu}$ operators are complicated linear combinations of $\psi_{e\mu} \sim \psi_{L\mu} + \psi_{R\mu}$ for $p = e$ and $\psi_{o\mu} \sim \psi_{L \mu} - \psi_{R \mu}$ for $p = o$ (see ref.~\onlinecite{Jones} for more details).   Here, $2D$ is the bandwidth cutoff, $\Lambda > 1$ is a numerical parameter (we use $\Lambda = 3$), and $\xi_n \sim 1$ is a dimensionless function of $\Lambda$.  The limit $\Lambda \to 1$ recovers the original model, albeit in a discrete basis.

To simulate the RG flow numerically, one introduces a series of dimensionless Hamiltonians by truncating the semi-infinite chains to $N+1$ sites each and setting the overall energy scale to be of order unity:
\bea
\nonumber H_N & = & \Lambda^{(N-1)/2} \left\{ \sum_{n=0}^{N-1} \sum_{p=e,o} \Lambda^{-\frac{n}{2}} \xi_n \left( f_{np\mu}^\dag f_{n+1,p \mu} + \textrm{h.c.} \right) \right. \\
\nonumber & & + \tilde{J} \sum_{p = e,o} f_{0 p \mu}^\dag \frac{\bs{\sigma}_{\mu \nu}}{2} f_{0 p \nu} \cdot \left( \bs{S}_L + \bs{S}_R \right) \\
\nonumber & & + \tilde{J} \left( f_{0 e \mu}^\dag \frac{\bs{\sigma}_{\mu \nu}}{2} f_{0 o \nu} + \textrm{h.c.} \right) \cdot \left( \bs{S}_L - \bs{S}_R \right) \\
\label{HN}
& & \left. + \tilde{K} \bs{S}_L \cdot \bs{S}_R + \tilde{V}_{LR} \left( f_{0e\mu}^\dag f_{0 e \mu} - f_{0 o \mu}^\dag f_{0 o \mu} \right) \right\}.
\eea
The parameters $\tilde{J}$, $\tilde{K}$, and $\tilde{V}_{LR}$ are rescaled versions of $J$, $K$, and $V_{LR}$ such that
\be
\label{hnlimit}
\frac{H}{D} = \lim_{N \to \infty} \frac{1}{2} \left(1 + \Lambda^{-1} \right) \Lambda^{-(N-1)/2} H_N.
\ee

The RG transformation is realized by numerically diagonalizing the series of Hamiltonians $H_N$, starting with $N=0$, using the lowest~\cite{footnote01} eigenvalues (and associated eigenvectors) of $H_N$ to define the $H_{N+1}$ matrix, until the energy spectrum ceases to change from one iteration to the next~\cite{footnote02}.   The energy scale of the spectrum of the dimensionless $H_N$ is of order unity, meaning that it describes $H$ at a temperature scale $T_N$ given by eq.~(\ref{hnlimit}) to be
\be
\label{tn}
T_N \sim \frac{1}{2} \left( 1 + \Lambda^{-1} \right) \Lambda^{-(N-1)/2} D.
\ee
This identification will allow us to use the flow of the NRG energy levels to measure the energy scale $T^*$.  Herein, we will take $D=1$ and measure all energy quantities in units of $D$.

\begin{figure}
\begin{center}
\includegraphics[width=0.48\textwidth,clip=true]{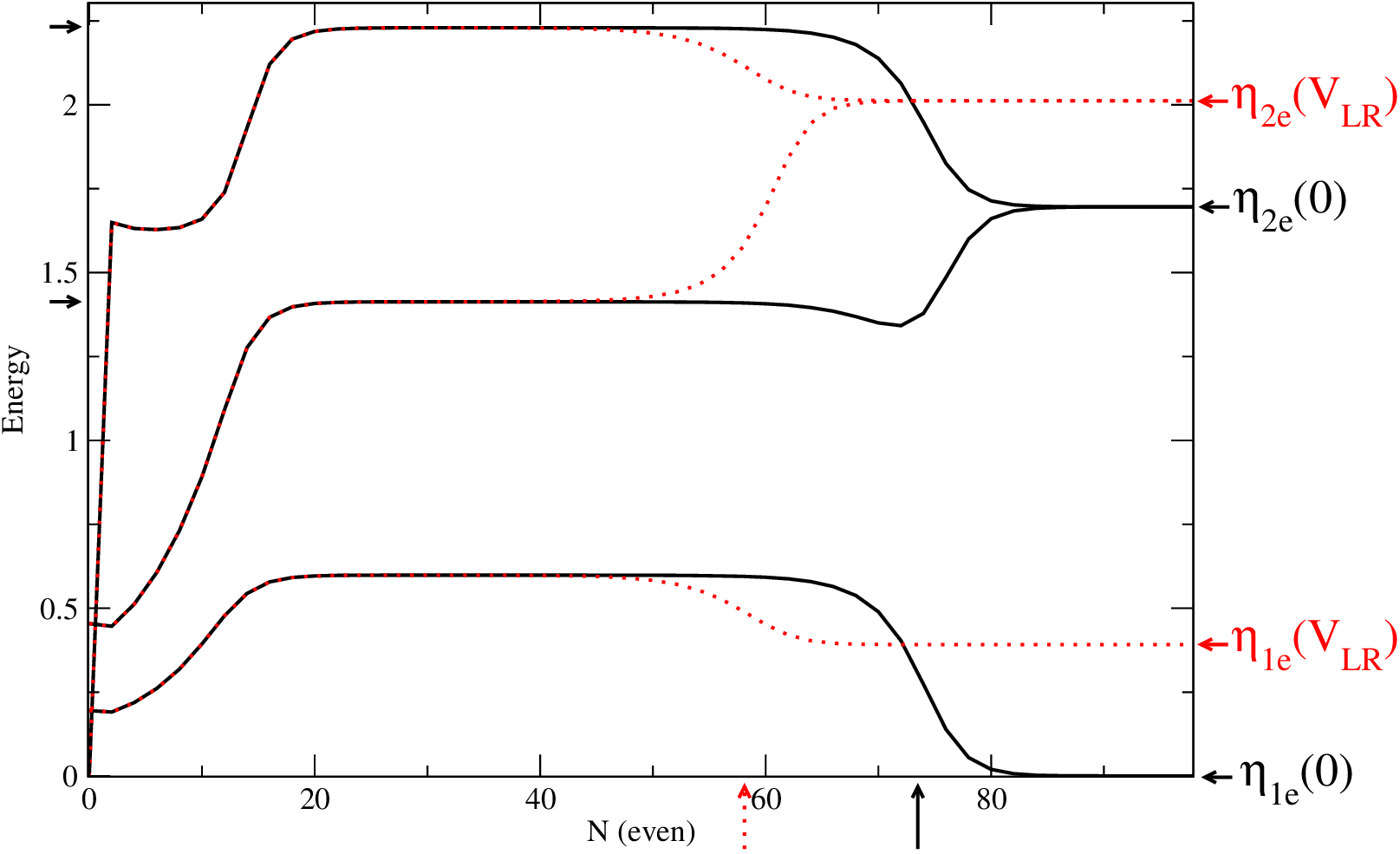}
\end{center}
\caption{\label{energylevels} The lowest three NRG energy levels in the charge-+1, spin-1/2, parity-even subspace as a function of even iteration parameter $N$.  A value of $\nu V_{LR} = 0$ was used to determine the solid black lines whereas a value of $\nu V_{LR} = 3.372 \times 10^{-7}$ was used for the red dotted lines.  The arrows on the left axis label the prediction of the second and third excitation of the QCP from ref.~\onlinecite{Affleck1995} (the first excitation is used to match the overall scale of the predicted spectrum with that from the NRG).  The arrows on the right axis label the first and second single-particle excitations used in the determination of the even channel phase shift, eq.~(\ref{NRGphase_shift}).  The arrows on the bottom axis label the value of $\langle N^* \rangle$ used to estimate $T^*$ for each of the two spectra.  Here, $\nu J = 0.15$ and $K = 1.30096478 \times 10^{-3} > K_c$ so that the final, stable fixed point is that of the LSP.}
\end{figure}
To see how this works in practice, consider the flow of energy levels when the Hamiltonian parameters have been tuned close to the quantum critical point.  A few such energy levels have been plotted in Fig.~\ref{energylevels}.  The energy levels flow close to those of the unstable QCP but eventually flow to a stable fixed point (for $V_{LR}=0$ it is either the LSP or KSP depending on the initial value of $K$; see black solid lines) at an iteration number $N^*$.     We identify the KSP and the LSP by looking at the quantum numbers and degeneracy of the energy levels in each fixed point regime and comparing to those expected from the Wilson chain forms of the fixed point Hamiltonians~\cite{Jones}.   This is not possible for the QCP which does not allow a simple Hamiltonian form.  Identifying the spectrum of the QCP requires more sophisticated techniques~\cite{Affleck1995}.

Quantitatively, we measure the value of $N_i^*$ when the $i^\mathrm{th}$ energy level crosses over from its QCP value to its stable fixed point value, then take the average $\langle N_i^* \rangle$ of the values $N_i^*$ for the first 20 energy levels.  The resultant value of $T^*$ is determined from eq.~(\ref{tn}) to be
\be
T^* = \frac{1}{2} \left(1+\Lambda^{-1}\right) \Lambda^{-( \langle N_i^* \rangle -1)/2}.
\ee

\begin{figure}
\begin{center}
\includegraphics[width=0.48\textwidth,clip=true]{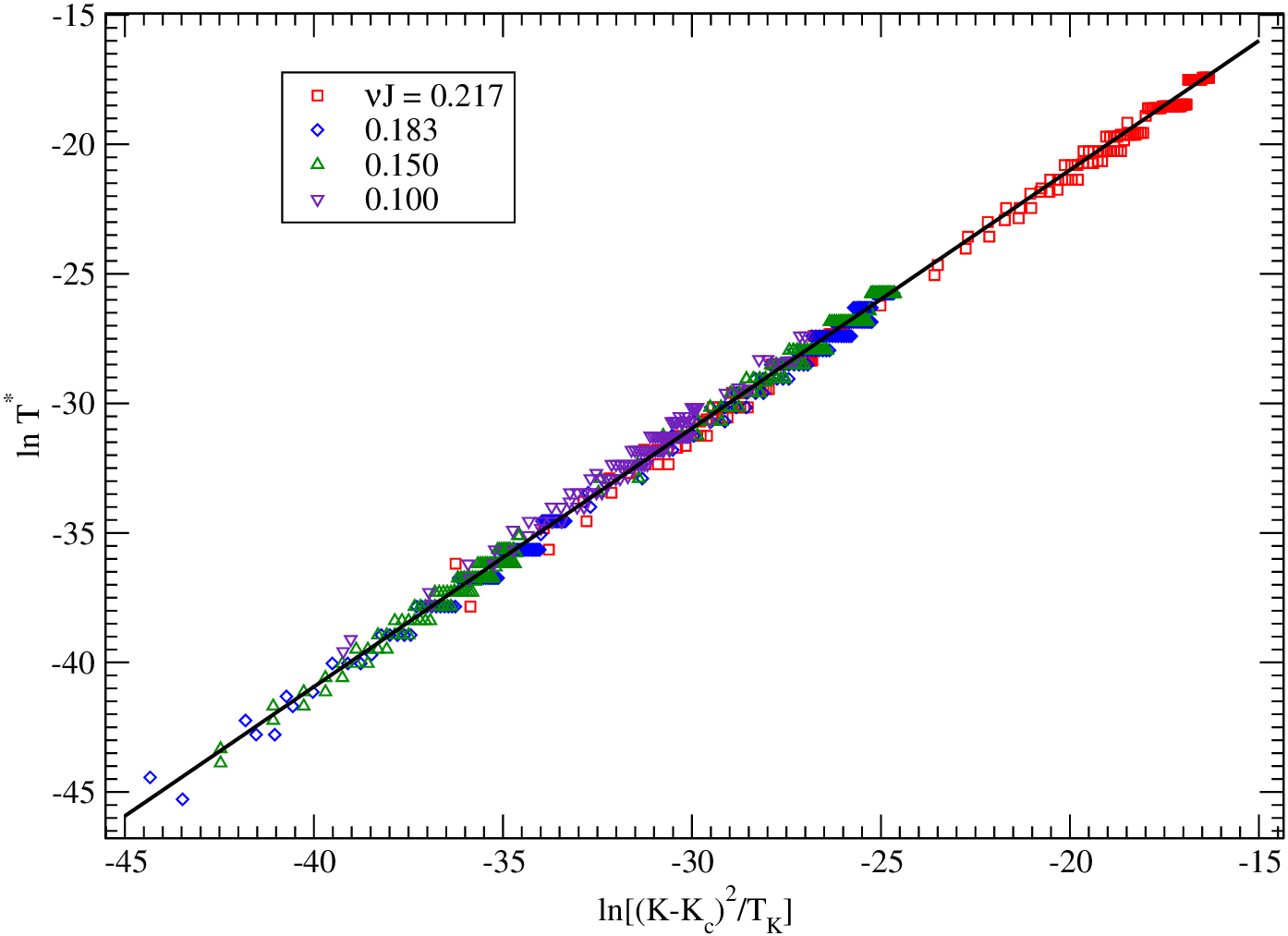}
\end{center}
\caption{\label{tstar_vs_k-kc} Data for $T^*$ as determined from the NRG with $V_{LR} = 0$.  The solid line is the best linear fit to all of the data.}
\end{figure}

To estimate the value of $a$ in eq.~(\ref{tstar}), we set $V_{LR}=0$ so that $T^* = T_{\delta K}$ and choose a value of $\nu J$.  We then obtain NRG spectra for a series of Hamiltonians with differing values of $K$, starting with~\cite{footnote04} $K = K_c$  then tuning $K$ away from $K_c$ until $T^* \sim T_K$ (\textit{i.e.}\ when the QCP is no longer reached in the RG).  From these spectra, $T^*$ is extracted as described above and a plot is made of $\ln T^*$ vs.\ $\ln \left[ \left( K - K_c \right)^2 / T_K \right]$;  see figure \ref{tstar_vs_k-kc}.  A linear fit is made to the data and compared with the expectation from eq.~(\ref{tstar}):
\be
\label{tstarvsk}
\ln T^* = m_a \ln \frac{ \left(K-K_c \right)^2}{T_K} + \ln a , \quad V_{LR} = 0
\ee
where $T_K$ is determined from the input value of $\nu J$ using eq.~(\ref{Tk}).  Obtaining the slope $m_a = 1$ from the fit provides a check on the dependence of $T^*$ on $K-K_c$ while the intercept of the fit provides an estimate of the dimensionless coefficient $a$.

\begin{figure}
\begin{center}
\includegraphics[width=0.48\textwidth,clip=true]{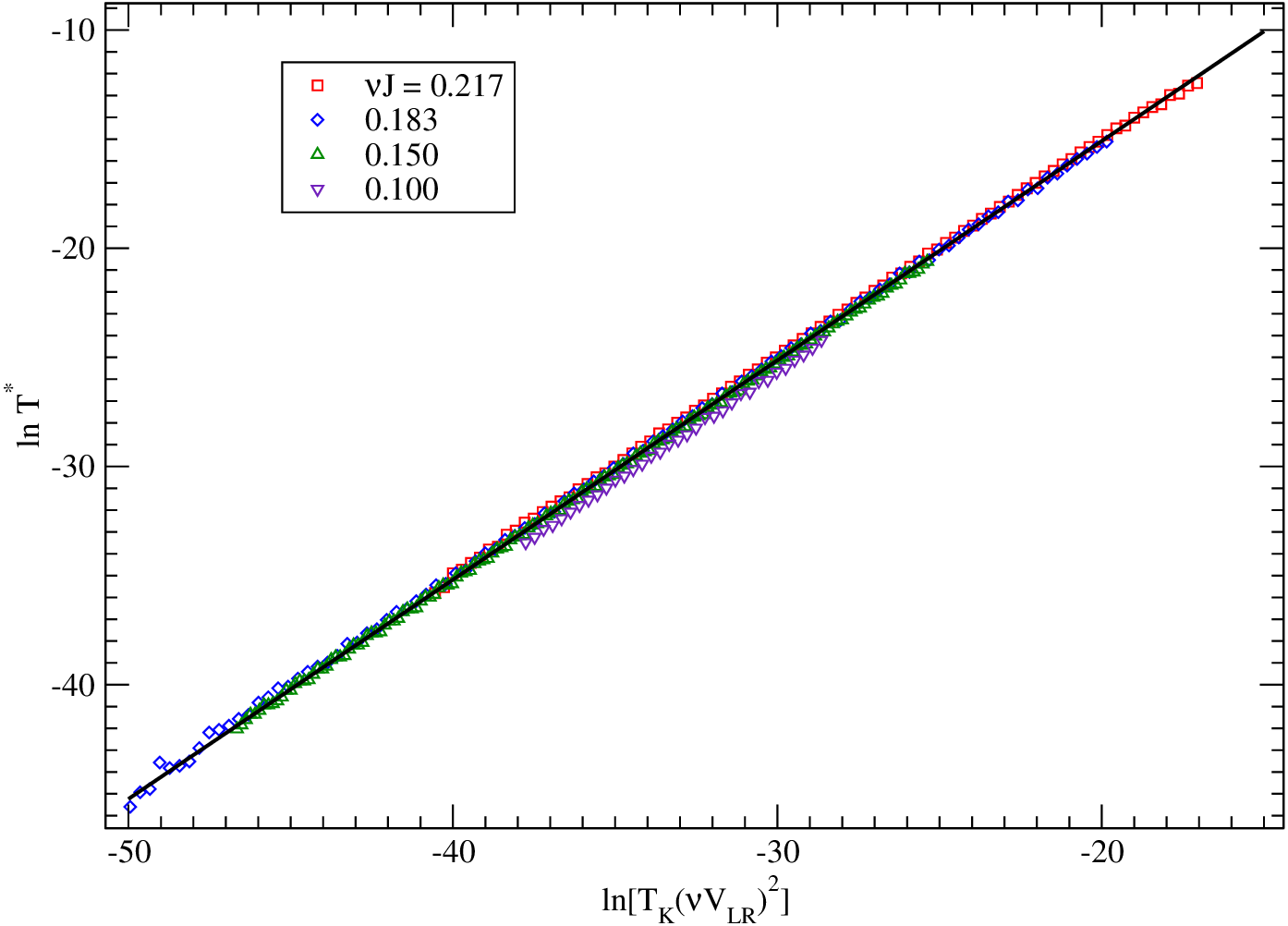}
\end{center}
\caption{\label{tstar_vs_V} Data for $T^*$ as determined from the NRG with $K = K_c$.  The solid line is the best linear fit to all of the data. }
\end{figure}

A similar procedure is used to measure $b$.  With $K=K_c$ so that $T^* = T_{LR}$, NRG data is obtained for a series of values of $V_{LR}$, starting with $V_{LR}=0$ and increasing $V_{LR}$ until $T^* \sim T_K$.  A plot is made of $\ln T^*$ vs.\ $\ln \left[ T_K (\nu V_{LR})^2 \right]$; see figure~\ref{tstar_vs_V}.  Again, a linear fit is made to the data and compared with the expectation from eq.~(\ref{tstar}):
\be
\label{tstarvsv}
\ln T^* = m_b \ln \left[ T_K (\nu V_{LR})^2 \right] + \ln b, \quad K = K_c.
\ee
As before, obtaining the slope $m_b = 1$ provides a check on the dependence of $T^*$ on $\nu V_{LR}$ and the intercept determines the value of $b$.  Consistent values of $a$ and $b$ over several values of $\nu J$ provide confirmation of the proposed crossover energy scale~(\ref{tstar}).

We have carried out the procedure described above for four values of $\nu J$.  The parameters $a$ and $b$ are tabulated in Table~\ref{resultstable} together with the error arising from the fit to the data.  In Figures~\ref{tstar_vs_k-kc} and~\ref{tstar_vs_V} we have plotted the data for all four of these iterations and find reasonably good data collapse.
\begin{table*}
\begin{center}
\begin{tabular}{|@{$\quad$}r@{$\quad$}r@{$\quad$}r@{$\quad$}r@{$\quad$}r@{$\quad$}r@{$\quad$}r@{$\quad$}r@{$\quad$}r@{$\quad$}r@{$\quad$}r@{$\quad$}|}
\hline
$\nu J$ & $T_K$ $(\times 10^{-3})$ & $K_c$ $(\times 10^{-3})$ & $m_a$ & $\Delta m_a$ & $a$ & $\Delta a$ & $m_b$ & $\Delta m_b$ & $b$ & $\Delta b$ \\
\hline
0.217 & 4.644 & 15.6194231  & 1.004 & 0.004  & 0.35 & 0.04 &  0.994 & 0.001  & 130 & 6 \\
0.183 & 1.811 & 5.55847415  & 0.999 & 0.005  & 0.33 & 0.05  &  0.997 & 0.002  & 125 & 8 \\
0.150 & 0.4929 & 1.30096469  & 1.001 & 0.004  & 0.37 & 0.04 &  0.999 & 0.001  & 114 & 3 \\
0.100 & 0.01436 & 0.02617379 & 0.99 & 0.01  & 0.4 & 0.2 &  1.009 & 0.004  & 100 & 14 \\
\hline
\end{tabular}
\end{center}
\caption{NRG results for four values of $\nu J$.  Note that all temperatures are measured in units of $D$ with $2D$ being the bandwidth in the leads.  The dimensionless parameters $a$ and $b$ are defined in eq.~(\ref{tstar}) and $\Delta x$ indicates the error in $x$ arising from the linear regression.} \label{resultstable}
\end{table*}

Before describing our analysis of $T^*$, it is interesting to note that, although we find that $K_c \sim T_K$ as is often quoted in the literature, we do not find the consistent value of $K_c = 2.2 T_K$ as reported in ref.~\onlinecite{Jones}.  Rather, we find $K_c = \alpha( \nu J ) T_K$ where $\alpha$ ranges monotonically (at least according to the four values obtained and listed in Table~\ref{resultstable}) from $\alpha (0.217) = 3.36$ to $\alpha(0.100) = 1.8$.   Although this does not change our conclusions regarding $T^*$, we simply point out that the relation between $K_c$ and $T_K$ may not be as simple as first presented~\cite{Jones}.

\begin{figure}
\begin{center}
\includegraphics[width=0.48\textwidth,clip=true]{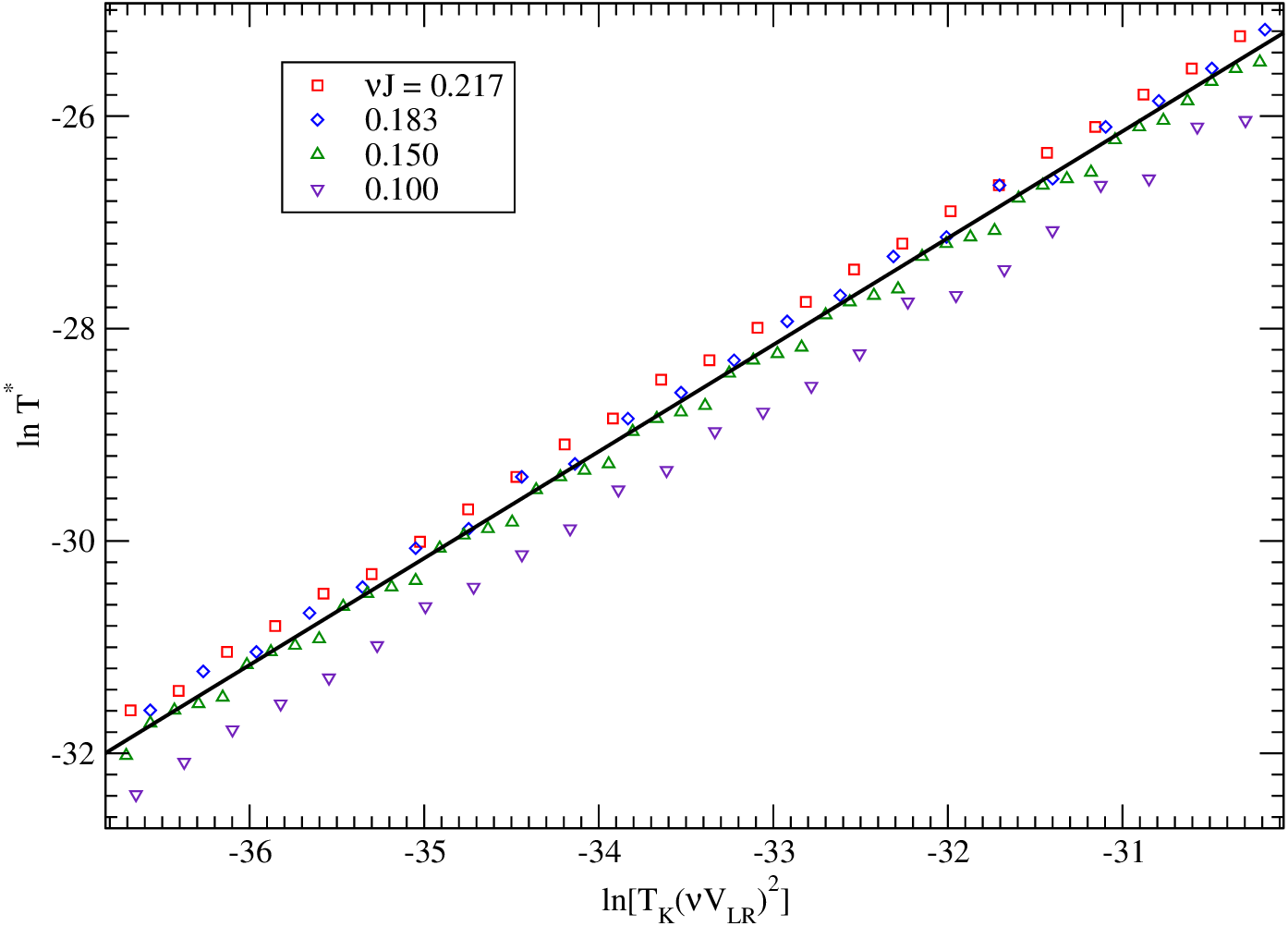}
\end{center}
\caption{\label{tstar_vs_V_closeup} An enlarged version of a section of Figure~\ref{tstar_vs_V}.}
\end{figure}

Returning to our analysis of $T^*$, it is seen that, while the value of $a$ is roughly of order unity, the value of $b$ here is two orders of magnitude larger than unity.  Furthermore, while the values of $a$ appear to be consistent for all values of $\nu J$, there seems to be a slight trend of $b$ decreasing with decreasing $J$.  This can be seen in Figure~\ref{tstar_vs_V_closeup} which is simply an enlarged area of Figure~\ref{tstar_vs_V}.  The fact that we consistently obtain $m_b \approx 1$ means that the $T^*$ dependence on $\nu V_{LR}$ is certainly quadratic but that the $J$ dependence of the coefficient in the second term of eq.~(\ref{tstar}) may be different than that included in the factor of $T_K$.

\begin{figure}
\begin{center}
\includegraphics[width=0.48\textwidth,clip=true]{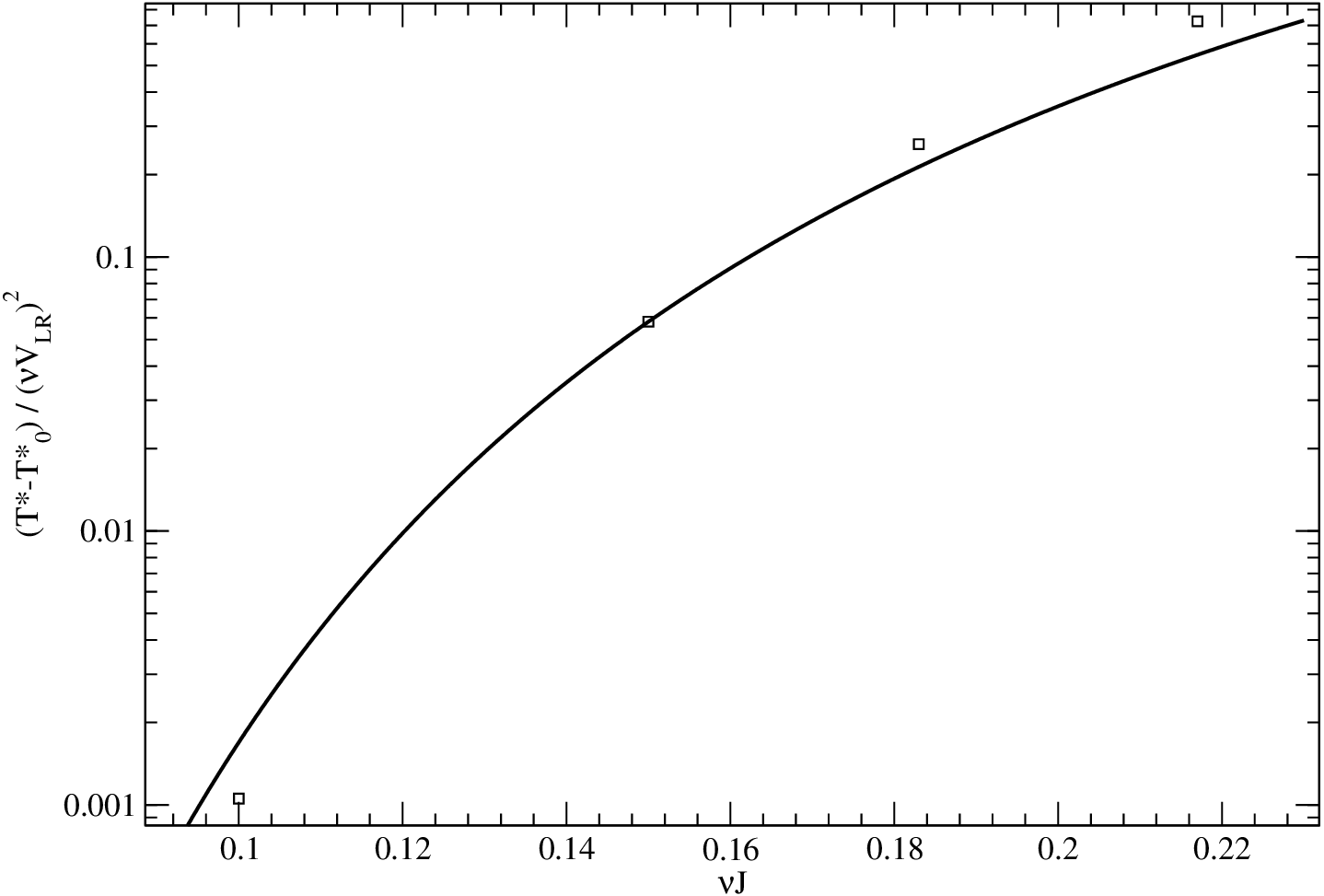}
\end{center}
\caption{\label{tstar-div-V2} The value of $T^*  / (\nu V_{LR})^2$ extrapolated to $\nu V_{LR} = 0$ for each of the four data sets (points).   Here, $K = K_c$.  A function of the form of eq.~(\ref{bfit}) is fit to the data (solid line) with the free parameter found to be $b = 118$.
}
\end{figure}

For further analysis, we look at the explicit $J$ dependence of the coefficient of $(\nu V_{LR})^2$ in eq.~(\ref{tstar}) to see how well it matches that of the predicted $b T_K$.  To do this, we set $K = K_c$ and plot in Figure~\ref{tstar-div-V2} the value of $T^* / (\nu V_{LR})^2$ extrapolated to $\nu V_{LR} = 0$ for each of the four data sets versus the corresponding value of $\nu J$.~\cite{footnote03}  To this data we have fit a function of the form
\be
\label{bfit}
\frac{T^*}{\left( \nu V_{LR} \right)^2} = b T_K = b \sqrt{\nu J} e^{-\frac{1}{\nu J}}
\ee
and obtained a value of $b = 118$.  This value is close to the mean of the four values of $b$ listed in Table~\ref{resultstable} and the above function provides a reasonable fit to the data.  However, as before, there is a hint of further $J$ dependence as the function appears to overestimate the $\nu J = 0.1$ data while underestimating the data at $\nu J = 0.217$.

\section{Fermi liquid theory}
\label{sec:FL}

Having determined the two energy scales $T_K$ and $T^*$, the latter with aid from the NRG, one can perform perturbative calculations in wide temperature ranges using effective theories appropriate for each regime. In the high temperature weak coupling regime $T \gg T_K$, one can apply perturbation theory in $J$ to the original model described by the Hamiltonian of eq. (1). In the intermediate temperature QCP, $T^* \ll T \ll T_K$, one should use the conformal field theory~\cite{Affleck92} (or the corresponding theory in the language of abelian
bosonization~\cite{Gan95}) and apply perturbation theory in the irrelevant operator. In the low temperature Fermi liquid (FL) phase, $T \ll T^*$, one can use a Fermi liquid theory in terms of single fermion scattering states as briefly reported in ref.~[\onlinecite{Sela09Noise}].

In this section we shall give a derivation of the effective FL Hamiltonian (see eq.~(\ref{Hjj})).  In sub-section A we derive the phase shift which characterizes the
Fermi liquid fixed point, and which depends on the ratio $(K-K_c)/[T_K\nu V_{LR}]$.
This is done using only the formula for the $T=0$ conductance on the line
of Fermi liquid fixed points derived in ref. [\onlinecite{Sela2009}], a standard
Fermi liquid formula for the conductance in terms of the phase shift, and a
symmetry argument. In sub-section B, we derive the leading
irrelevant interactions and corresponding coupling constants on the
line of Fermi liquid fixed points.
This is done starting with the abelian
bosonization description of the QCP~\cite{Gan95,Sela2009} which uses Majorana fermions that are non-locally related to the original Dirac fermion fields in eq.~(\ref{2IKH}).  We will then see that, at temperatures lower than $T^*$ and with the relevant operator present,  single electron scattering states become the correct particles in terms of which the FL theory is conveniently written.

\subsection{Phase Shift Analysis}
\label{sec:FLPS}

Before describing the derivation of the FL Hamiltonian, we present a simple derivation of the form of the phase shifts in terms of which the single electron scattering states are defined.  These are then compared to the phase shifts extracted from the fixed point NRG spectrum.  Since one can derive the same form for these phase shifts from the fixed point analysis described at the end of this section, numerical confirmation of the phases shifts provides additional support for the analytic calculation of the FL Hamiltonian.

In the simpler FL theory of the single channel Kondo effect\cite{Nozieres74}, the Hamiltonian is written in terms of weakly interacting fermionic scattering states which are simply the original Dirac electrons in which the zero temperature scattering phase shift is incorporated.  The same holds true for the two-impurity model under consideration.   Using the $L \leftrightarrow R$ symmetry, the original Dirac fermions satisfy a FL boundary condition (BC)
\bea
(\psi_{L \mu} + \psi_{R \mu})(0^+) = e^{2 i \delta_e}(\psi_{L \mu} + \psi_{R \mu})(0^-), \nonumber \\
\label{ps_evenodd}
(\psi_{L \mu} - \psi_{R \mu})(0^+) = e^{2 i \delta_o}(\psi_{L \mu} - \psi_{R \mu})(0^-).
\eea
Furthermore, using the special particle-hole symmetry
\be
\psi_{L \mu} \to \psi_{L \mu}^\dagger,~~~\psi_{R \mu} \to - \psi_{R \mu}^\dagger,
\ee
it follows~\cite{Affleck1995} that $\delta_e = - \delta_o \equiv \delta$.

To calculate the phase shift $\delta$, we use its relation to the zero temperature conductance~\cite{Georges1999}
\be
\frac{h}{2 e^2} G = \sin^2 \left( \delta_e - \delta_o \right) = \sin^2 2 \delta.
\ee
Comparing this with\cite{Sela2009}
\be G  =  \frac{2 e^2}{h} \frac{T_{LR}}{T^*} \label{GT0}\ee
we can immediately extract the form of $\delta$
\bea
\nonumber
\delta & = &\frac{1}{2} \arg (\sqrt{T_{\delta K}} + i \sqrt{T_{LR}}) \\
\label{ps_analytic}
& = & \frac{1}{2} {\rm{arg}}\left(\sqrt{\frac{a}{b}} \frac{K-K_c}{T_K} + i \nu V_{LR} \right)
\eea
where, in the last equality, we have substituted the expressions for $T_{\delta K}$ and $T_{LR}$ from eq.~(\ref{TdKTLR}).

For $V_{LR}>0$, the phase shift $\delta$ changes from $0$ to $\pi/2$
as function of $K$, and it takes the value of $\pi/4$ at $K=K_c$.
This agrees with the numerical results of Jones \textit{et.~al.}~\cite{Jones}
While the original electrons suffer a phase shift, the
single particle scattering
states $\Psi_{j \mu}$ incoming from lead $j =1,2=
(L,R)$, defined by
\bea
\label{eq:scattering}
 \Psi_{j \mu}(x) &=& \theta(x) \psi_{j \mu}(x)+\sum_{j'} \theta(-x) s_{j j'} \psi_{j'
 \mu}(x), \nonumber  \\
s&=& \begin{pmatrix} \cos 2 \delta&- i \sin 2 \delta\\ - i \sin 2 \delta& \cos 2 \delta  \end{pmatrix},
\eea
are continuous at the origin: $\Psi_{i \mu}(0^+) = \Psi_{i
\mu}(0^-)$.

The form of this predicted phase shift, eq.~(\ref{ps_analytic}), can be compared with the phase shift derived from the NRG fixed point spectrum.  To extract the phase shift, one looks at the many-body NRG energy spectrum for the Hamiltonian $H_N$ of eq.~(\ref{HN}) where $N$ is large enough such that the RG has reached one of the FL fixed points.  Unlike the QCP, the LSP and KSP many-body spectra are made up of two channels of single-particle/hole excitations combined in such a way so as to respect Fermi statistics.  By looking at the quantum numbers of the lowest many-body NRG energy levels, one can determine the two lowest single-particle/hole excitations $\eta_{1p}$ and $\eta_{2p}$ in each of the $p = e$ (even), $o$ (odd) channels.  From these we define the phase shift as
\be
\label{NRGphase_shift}
\delta_p = Q_{1p}\frac{\eta_{1p}}{\eta_{1p} + \eta_{2p}}  \pi
\ee
for the case of even $N$.  Here, $Q_{1p} = \pm 1$ is the charge of the lowest excitation in the $p$ channel indicating whether the spectrum is shifted up ($Q_{1p} = 1$, $\delta_p > 0$) or down ($Q_{1p} = -1$, $\delta_p < 0$) relative to the spectrum with $V_{LR} = 0$. The $\eta$'s are marked in figure~\ref{energylevels}.  The definition for odd $N$ is simply shifted by $\pi / 2$.   This definition of the phase shift at the fixed point follows closely that used in ref.~\onlinecite{Malecki2010} for a model of a quantum dot in an Aharonov-Bohm ring.

\begin{figure}
\begin{center}
\includegraphics[width=0.48\textwidth,clip=true]{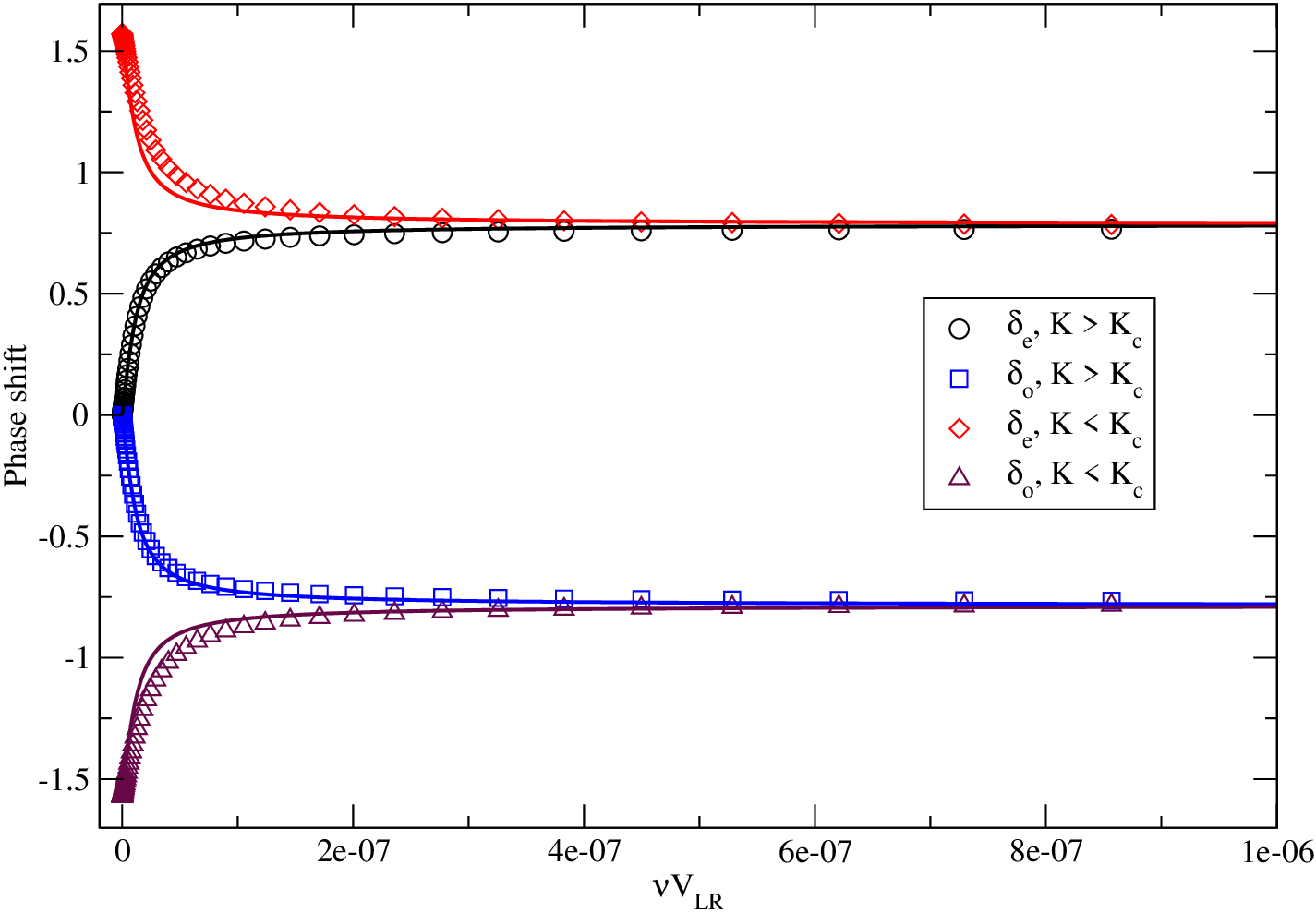}
\end{center}
\caption{\label{fig:phase_shifts} The phase shifts $\delta_e$ and $\delta_o$ in the even and odd channels (respectively) of the FL fixed point as derived from the NRG (points) with $\nu J = 0.15$ as well as those phase shifts predicted analytically by eqs.~(\ref{ps_evenodd}) and~(\ref{ps_analytic}) (lines).  The values for $T_K$, $K_c$, $a$, and $b$ appearing in eq.~(\ref{ps_analytic}) are taken from the $\nu J = 0.15$ line of Table~\ref{resultstable}.  We have done this for both the LSP, $K = 1.30096478 \times 10^{-3} > K_c$, and the KSP,  $K = 1.30096452 \times 10^{-3} < K_c$).  }
\end{figure}
In Figure~\ref{fig:phase_shifts}, we have plotted the phase shifts as derived in this way from the NRG and compared them with those predicted from eq.~(\ref{ps_analytic}) for both $K > K_c$ and $K < K_c$.  We find the agreement to be quite good and take this as support for our analysis of the FL fixed point.

Slight deviations from the continuum model used to derive eq.~(\ref{ps_analytic}) are known to exist due to the fact that $\Lambda > 1$ in the discrete NRG Hamiltonian (see ref.~\onlinecite{Malecki2010} for a discussion of this effect).  Furthermore, we obtain better agreement with the NRG for the phase shifts in the LSP than we do for phase shifts in the KSP.  The nature of this discrepancy looks very similar to that due to the presence of an additional potential scattering term that is generated by the Kondo interaction in the screening channel when particle-hole symmetry is broken~\cite{Malecki2010, Cragg1978}.  Since the presence of a non-zero $V_{LR}$ breaks particle hole symmetry, one would expect such an effect but only in the KSP where Kondo screening occurs.  This is precisely what is seen in Figure~\ref{fig:phase_shifts}.  However, the form of this additional potential scattering was only derived for single-channel, single-impurity models~\cite{Malecki2010, Cragg1978} so more analysis is required to determine for certain if this is the nature of the discrepancy in the KSP phase shifts.  Nevertheless, it is clear that eq.~(\ref{ps_analytic}) captures the leading order contribution to the phase shift for the entire manifold of fixed points.

\subsection{Fermi Liquid Hamiltonian}

We now turn our attention to the derivation of the FL Hamiltonian.  Using abelian bosonization, one can write the original free Fermion
theory (eq.~(\ref{2IKH}) with $J \to 0$ and $V_{LR} \to 0$) in terms of 8 chiral Majorana
Fermions $\chi_i ^A$ associated with the real ($\chi_1^A=\frac{\psi_A^\dagger+\psi_A}{\sqrt{2}}$) and imaginary
($\chi_2^A=\frac{\psi_A^\dagger-\psi_A}{\sqrt{2}i}$) parts of the charge, spin, flavor and spin-flavor fermions $\psi_A \sim e^{-i \phi_A}$,
$(A = c,s,f,X)$. The bosonic fields $\phi_A$ are linear combinations of the four bosonic fields associated with the original Dirac fermions, $\psi_{i \mu} \sim e^{-i \phi_{i \mu}}$, ($i=L,R=1,2,~\mu=\uparrow,\downarrow=1,2$) given by $\{ \phi_c ,\phi_s,\phi_f,\phi_X \} = \frac{1}{2} \sum_{i \mu}\phi_{i \mu} \{1,(-1)^{\mu+1},(-1)^{i+1},(-1)^{\mu+i} \}$.

The free Hamiltonian is $H_0[\{\chi'\}] =
\frac{i}{2} \sum_{j=1}^8\int dx \chi'_j
\partial_x \chi'_j$ where $\{\chi'_1,...,\chi'_8\}=\{  \chi_2^X,\chi_1^f,\chi_2^f,\chi_1^X,\chi_1^c,\chi_2^c,\chi_1^s,\chi_2^s
\}$ is an arbitrary relabeling of the 8 fields.  Turning on the Kondo coupling $J$ at $V_{LR}=0$, the QCP is obtained at $K=K_c$. It is described simply in terms of a change in the BC relative to the free
case in which $\chi_i'(0^-) =  \chi_i'(0^+)$, ($i=1,...8$). The change in BC occurs only for the
first Majorana fermion, $\chi_1'(0^-) = - \chi_1'(0^+)$. In our chirality convention for the 1 dimensional
fields, the region $x>0$ ($x<0$) corresponds to the
incoming (outgoing) part of the field.

For energies $\ll T_K$, we define a new basis
\bea
\label{chi}
\chi_1(x) &=& \chi_1'(x)~ {\rm{sgn}}(x),\nonumber \\
\chi_i (x)&=& \chi'_i(x),~~~(i=2,\ldots,8)
\eea
and write the leading terms in the Hamiltonian describing
deviations from the QCP due to finite $K-K_c$ as well as finite $V_{LR}$ as $H_{QCP}= H_0[\{\chi\}] +
\delta H_{QCP}$ with~\cite{Sela2009}
\begin{eqnarray}
\label{eq:HNFL} \delta H_{QCP}=i\sum_{i=1}^2  \lambda_i\chi_i(0) a.
\end{eqnarray}
Here, $a$ is a local Majorana fermion, $a^2=1/2$. Both terms in $\delta
H_{QCP}$ have critical dimension $1/2$ so that they destabilize the
QCP. The coupling constants satisfy $\lambda_1 \propto K-K_c$, $\lambda_2 \propto V_{LR}$ and
\be
\label{Tstar}
T_{\delta K} = \lambda_1^2,~~~T_{LR}=\lambda_2^2,~~~T^*=\lambda_1^2+\lambda_2^2
\ee
where $T^*$ is also given in eq.~(\ref{tstar}) with the coefficients $a$ and $b$ determined numerically in Table 1. Below the crossover scale $T^*$, the
system flows to FL fixed points whose nature depend on the ratio
$\lambda_1/\lambda_2$.

The crucial observation is that only the linear combination
$[\lambda_1 \chi_1(x)+ \lambda_2 \chi_2(x)]/\lambda$ of the 8 Majorana
fermions at the quantum critical point participates in this crossover. Here $\lambda = \sqrt{\lambda_1^2 + \lambda_2^2}$. It can be shown
that the effect of $\delta H_{QCP}$ is to modify the BC for only this
linear combination by a simple sign change at the boundary, $(\lambda_1 \chi_1(0^+)+ \lambda_2 \chi_2(0^+))/\lambda$ = - $(\lambda_1 \chi_1(0^-)+ \lambda_2 \chi_2(0^-))/\lambda$. In order
to write down the FL fixed point Hamiltonian, we define a new basis
with modified BC, $\{\eta\}$, where
\bea
\label{eta}
\eta_1 (x)&=& {\rm{sgn}}(x) (\lambda_1 \chi_1(x)+
\lambda_2 \chi_2(x))/\lambda~, \nonumber \\
\eta_2 (x)&=& (-\lambda_2
\chi_1(x)+ \lambda_1 \chi_2(x))/\lambda, \nonumber \\
\label{etadef}
\eta_i(x) &=& \chi_i(x),~~~(i=3,\ldots,8).
\eea
We can therefore write the Hamiltonian for the FL
fixed points as $H_{FL} = H_0[\{\eta\}]+\delta H_{FL}$. For clarity, we recapitulate the notation: in the weak coupling regime we defined the set of Majorana fermions $\{\chi' \}$;  in the QCP, the Hamiltonian eq.~(\ref{eq:HNFL}) is written in terms of $\{\chi \}$ given by the relation eq.~(\ref{chi}); in the FL regime, the Hamiltonian will be written in terms of the $\{ \eta \}$ operators defined in eq.~(\ref{etadef}).

Near a FL fixed point the leading interactions
have scaling dimension 2, which lead to corrections of various
quantities which vanish at zero temperature as $T^2$. In our local
theory, the interaction $\delta H_{FL}$ acts only at
$x=0$ and involves uniquely $\eta_1$ which is the only field
participating in the crossover in eq.~(\ref{eq:HNFL}). The only
possible such operator is
\begin{equation}
\label{eq:FL8} \delta H_{FL} =\lambda_{FL} i   \eta_1
\partial_x \eta_1 |_{x=0}.
\end{equation}
In the FL theory, $T^*$ acts as a high energy cutoff. From dimensional analysis, the coupling constant scales as $\lambda_{FL}\sim
\frac{1}{T^*}$.  Since eq.~(\ref{eq:HNFL}) is quadratic, we
can determine the coefficient exactly by matching at low temperatures the results of calculations of a physical quantity (\textit{e.g.}\ the conductance of a quantum dot~\cite{Sela2009}) as calculated either using eq.~(\ref{eq:HNFL}) or eq.~(\ref{eq:FL8}).  In this way, we obtain $\lambda_{FL} =\frac{4}{T^*}$.

We have derived the FL scattering states $\Psi_{j \mu}$, $j = 1, 2 = L, R$, in eq.~(\ref{eq:scattering}) and now rewrite the FL interaction of eq.~(\ref{eq:FL8}) in terms of them.  First we write $\delta H_{FL}$ using eqs. (\ref{chi}) and (\ref{eta}) as
\bea
\label{h1}
\delta H_{FL} &=& \frac{4 i }{T^*} [ \cos^2 (2 \delta) (\chi_1' i \partial_x \chi_1') + \sin^2 (2 \delta) (\chi_2' i \partial_x \chi_2') \nonumber \\
&+&{\rm{sgn}}(x) \frac{1}{2} \sin (4 \delta)  (\chi_1' i \partial_x \chi_2'+\chi_2' i \partial_x \chi_1') ].
\eea
These quadratic operators of dimension 4 can be written as a product of two normal ordered quadratic forms using
\bea
\label{h2}
\chi_1' i \partial_x \chi_1'+\chi_2' i \partial_x \chi_2' &=& (i :\chi_1' \chi_2':)^2, \nonumber \\
\chi_1' i \partial_x \chi_1'-\chi_2' i \partial_x \chi_2' &=& (i :\chi_1' \chi_3':)^2 -(i :\chi_2' \chi_3':)^2\nonumber, \\
\chi_1' i \partial_x \chi_2'+\chi_2' i \partial_x \chi_1' &=& \{ :\chi_1' \chi_3':,:\chi_3' \chi_2': \}_+,
\eea
where $\{, \}_+$ stands for the anticommutator.

These normal ordered quadratic forms are related to the flavor currents: \be
\vec{J}_f =\sum_{\mu , i , j} :\Psi^\dagger_{i \mu} \frac{\vec{\tau}_{ij}}{2} \Psi_{j \mu}:
\ee
where $\vec{\tau}_{ij}$ are Pauli matrices.  Indeed in the range $x>0$ at which the scattering states coincide with the original fermions $\Psi(x>0) = \psi(x>0)$, these can be expressed as
\be
\label{jchichi}
J_f^x=i : \chi_1' \chi_2':,~~~J_f^y=i :\chi_1' \chi_3':,~~~J_f^z=i :\chi_3' \chi_2':.
\ee
Using eqs.~(\ref{h1}), (\ref{h2}) and (\ref{jchichi}), we finally obtain
\bea
\label{Hjj}
\delta H_{FL} = \frac{2}{T^*}  \vec{J}_f^T \hat{M} \vec{J}_f,
\eea
where
\be
\hat{M}= \begin{pmatrix} 1 &0&0\\ 0 & \cos 4 \delta & -\sin 4 \delta \\0& -\sin 4 \delta & -\cos 4 \delta  \end{pmatrix} . \nonumber
\ee
One can see that, due to the potential scattering term $\propto V_{LR}  J_f^x$, the flavor SU(2) symmetry is reduced down to U(1). One can rewrite $\delta H_{FL}$ in different ways, for example, in ref.~\onlinecite{Sela2009}, the same equation is written in an explictly spin SU(2) symmetric way.

We emphasize the universality of the derived FL Hamiltonian: for any value of the ratio of original parameters $V_{LR} / (K-K_c)$,  all coupling constants of $H_{FL}$ are determined up to the overall  energy scale $T^*$ which was calculated numerically here. The universality follows from strong restrictions due to a large symmetry that emerges close to the quantum critical point~\cite{Affleck92} and leads to
the simple form of $H_{QCP}$ in eq.~(\ref{eq:HNFL}). One can see
that $H_0^{QCP}$ has an SO(8) symmetry. Due to $\delta
H_{QCP}$, the crossover from the QCP to FL fixed points has an SO(7)
symmetry represented by rotations of the vector $(\chi_2
,\ldots,\chi_8)$.
This symmetry considerably restricts the possible interactions and
sets relations between the different coefficients in
eq.~(\ref{Hjj}). The conditions of validy of $\delta H_{FL}$ is small deviations from the QCP $|K-K_c| \ll T_K$, $\nu V_{LR} \ll 1$. In addition, a scale separation $T^* \ll T_K$ is required. This scale separation was shown to hold using our NRG calculations as discussed in detail in \S~\ref{sec:nrgtstar}.

In practice, the SO(8)
symmetry at the quantum critical point is broken by marginal and irrelevant operators
at the QCP such as the leading irrelevant operator $
 (\partial_x \chi_1)a$ at the quantum critical point.  However, these will be associated with
a small parameter $1/T_K$ and, hence, are neglected at $T^* \ll T_K$.
For finite $V_{LR}$ or $K-K_c$, additional marginal
and irrelevant terms are produced at the quantum critical point, part of which were
present before. However, close enough to the quantum critical point, namely for $\nu
V_{LR} \ll 1$ and $|K-K_c| \ll T_K$,
those perturbations can be safely ignored.

\section{The Two Impurity Anderson Model}
\label{sec:disc}
We have presented a detailed analytic and numerical calculation of the scale for the crossover away from the QCP in the two-impurity Kondo model.  It is at this scale that the system flows to a Fermi liquid fixed point on a one-dimensional manifold of fixed points parametrized by $V_{LR} / (K - K_c)$.  The Fermi liquid behavior along this manifold has been derived in detail and confirmed numerically.   Furthermore, the more precise calculation of $T^*$ presented in this paper provides a more accurate estimate of the linear conductance written explicitly in terms of the crossover scales $T^*$ and $T_{LR}$, which in the limit $T \ll T^*$ is given by~\cite{Sela2009}
\be
\label{G_Sela2009}
G  =  \frac{2 e^2}{h} \frac{T_{LR}}{T^*} \left[ 1 - \left( \frac{2 \pi T}{\sqrt{3} T^*} \right)^2 \right] .
\ee
As discussed in the previous section, the $T=0$ limit of this expression for the conductance agrees well with our NRG results.

We now discuss the applicability of our analysis to double quantum dot setups. First, real quantum dots include charge fluctuations and hence are described by the Anderson model rather than Kondo model. Second, in real quantum dots there are additional tunneling process not included in our model with the direct tunneling $V_{LR}$ term. In those additional terms the tunneling electron's spin interacts with the impurities. We will re-express our Kondo model results in terms of Anderson model
parameters.\cite{Sela2009} We define the Anderson model with
hopping $t_d$ between the conduction electrons and the impurities and hopping $t_{LR}$ between the two impurity sites on which there is a Coulomb repulsion $U$ and resonant energy level $\eps_d = -U/2$.
Carrying out a Schrieffer-Wolff calculation up to third order in the tunneling terms, the Kondo parameters are given by:
\bea J &\approx& 8 t_d^2 / U\nonumber \\
K&\approx& 4t_{LR}^2/U\nonumber \\
V_{LR} &\approx& 0 \nonumber \\
\label{andersonTK}
T_K &\approx& U \sqrt{\frac{\pi \nu J}{16}} \exp \left( {-\frac{1}{\nu J}} \right).
\eea
(See ref.~\onlinecite{Haldane1978} for the final formula in Eq. (\ref{andersonTK}) which is the same as that used in ref.~\onlinecite{Izumida2000}.)
Interestingly, the direct tunneling operator is completely absent at $\eps_d = -U/2$. Instead, one obtains the tunneling terms
\bea
\label{deltaHLR}
\delta H &=&  V_{LR}^K  (\psi^\dagger_{L \mu} \psi_{R \mu}+ \psi^\dagger_{R \mu} \psi_{L \mu}) \vec{S}_L \cdot \vec{S}_R \\
&+&V_{LR}^{DM}    (- i  \psi^\dagger_{L \mu} \vec{\sigma}_{\mu \mu'} \psi_{R \mu'}+i  \psi^\dagger_{R \mu} \vec{\sigma}_{\mu \mu'} \psi_{L \mu'})\cdot (\vec{S}_R \times \vec{S}_L), \nonumber
\eea
describing tunneling mediated by interaction with the impurity spins. Here $V_{LR}^K=V_{LR}^{DM}=\frac{16 t_d^2 t_{LR}}{U^2}$.

We now analyze these operators from a field theoretical perspective, assuming that originally the
coupling constants are very small, and check if they eventually destabilize the fixed point by turning on the relevant dimension $1/2$ operator  with coupling constant $\lambda_2$.  It is convenient to introduce three types of particle hole symmetry:
\bea
\rm{ph1} &:& \psi \to \sigma^y \psi^\dagger , \nonumber \\
\rm{ph2} &:& \psi \to \sigma^y \tau^z \psi^\dagger , \nonumber \\
\rm{ph3} &:& \psi \to \sigma^y \tau^x \psi^\dagger ,
\eea
where the first two were introduced in Ref.~\onlinecite{Affleck1995} [here $\vec{\sigma} (\vec{\tau})$ are Pauli matrices acting in spin (channel) space]. Both terms in $\delta H$ are odd under ph1 and even under ph2. The relevant tunneling operator with coefficient $\lambda_2$, has the same symmetry as $V_{LR}$,~\cite{Affleck1995} which is also odd and even with respect to ph1 and ph2, respectively. To distinguish between the two terms in $\delta H$ we introduce ph3. The operators with
coupling constants $V_{LR}$ and $V_{LR}^{K}$ are both odd under p-h3.
 The relevant tunneling operator with coefficient $\lambda_2$, has the same symmetry as $V_{LR}$,~\cite{Affleck1995} and  must therefore
also be odd under p-h3.  Hence $V_{LR}^K$ should generate $\lambda_2$. It can be checked that the $V_{LR}^{DM}$ operator is even under p-h3; hence it can not generate $\lambda_2$.  Since both coefficients $V_{LR}^{K}$ and $V_{LR}^{DM}$ are initially small, we keep only the former which will grow under renormalization group near the critical point.  Naively, we might expect that we could replace
$\vec{S}_L \cdot \vec{S}_R$ by its expectation value at the quantum critical point in calculating $\lambda_2$:
$\vec{S}_L \cdot \vec{S}_R  \approx -1/4$, corresponding to equal probability of the spins to be found in a singlet or in a triplet state.~\cite{Jones} Hence, $\delta H$ effectively becomes
\be
\label{naive}
\delta H \to - \frac{4 t_d^2 t_{LR}}{U^2}    (\psi^\dagger_{L \mu} \psi_{R \mu}+ \psi^\dagger_{R \mu} \psi_{L \mu}) .
 \ee Based on this na\"{i}ve order of magnitude estimate it is interesting to discuss how our results for $T^*$ affect the feasibility of observing the QCP in such a quantum dot experiment. The criterion for the observability of the QCP is the separation of the two crossovers, namely the crossover from weak coupling to QCP occuring at scale $T_K$ and the crossover from QCP to FL occuring at scale $T^*$. This requirement reads $T_K/T^* \gg 1$.
Using the Kondo-Anderson correspondence with Eq.~(\ref{naive}), the crossover temperature is:
\be
\label{tlr}
T_{LR} = \frac{b T_K (\nu J)^2}{4 U^2} t_{LR}^2.
\ee
Evaluated at $t_{LR} = t_{LR}^c \equiv \sqrt{UK_c/4}$ and estimating $K_c = 4 (t_{LR}^c)^2 / U \approx 3 T_K$ and $b \approx 115$, we obtain
\be
\label{TLRest}
\frac{T_{LR}}{T_K} \approx 20 \frac{ T_K (\nu J)^2}{U}.
\ee
In order to estimate the magnitude $T_K / T^*$, we use a typical value of $U = 1.5 \mathrm{meV}$ and a value of $\nu J = 0.217$.  Rather than estimating $T_K$ using the Kondo model formula~(\ref{Tk}), we use the Kondo temperature of eq.~(\ref{andersonTK}) as derived from an Anderson model~\cite{Haldane1978} which is more applicable to quantum dots.  In this case, we obtain $T_K \approx 3.1 \mu \mathrm{eV} \approx 0.04 \mathrm{K}$ so that the ratio of scales is $T_K / T^* \approx 500 \gg 1$.  Although it would be very difficult to attain temperatures $T < T_K$ for such a small value of the Kondo temperature, holding $U = 1.5 \mathrm{meV}$ fixed in eq.~(\ref{TLRest}) suggests that one can obtain a ratio of $T_K / T^* \approx 10$ at a Kondo temperature of $T_K \approx 40 \mu \mathrm{eV} \approx 0.5 \mathrm{K}$ ($\nu J \approx 0.4$), a temperature that can be obtained in modern quantum dot experiments. Hence,  we would conclude that it is possible to realize the QCP in a double
 quantum dot system.

However, NRG calculations~\cite{Sakai1992,Izumida2000,zitko} on the two impurity Anderson model show that  in order to achieve scale separation $T_K \gg T^*$, much smaller values of $\nu J  \propto \Gamma / U$ are needed in the Anderson model; whereas in the former paragraph it was estimated that $\nu J = 0.4$ would be sufficient to fulfill a scale separation, it appears that $\nu J$ should be at least one order of magnitude smaller, corresponding to $\Gamma / U =0.05$ in the Anderson model,~\cite{zitko} to achieve such scale separation.

This means that the operator $\delta H$ in Eq.~(\ref{deltaHLR}) turns out to lead to a much larger crossover energy scale $T^*$ as compared to the direct tunneling $V_{LR}$ term. Thus, we conclude that in order to observe the quantum critical point in quantum dots one should achieve the weak tunneling and very low temperature regime, which goes beyond the current ability. Otherwise, another parameter
must be fine-tuned in order to make $\lambda_2$ small.
Alternatively, more elaborate schemes may be used, where the two leads are separated by multiple quantum dots hence suppressing the tunneling.~\cite{Zarand2006} Further analysis is required  to explain why the operator $\delta H$ in Eq.~(\ref{deltaHLR}) above and the $V_{LR}$ term in Eq.~(1) give so different crossover energy scales.

\section*{Acknowledgements}

The authors would like to thank Barbara Jones for helpful discussions during the course of this research. We thank Rok \v{Z}itko for kindly showing us his unpublished results on the two impurity Anderson model. Shortly after we had obtained the tunneling operator $\delta H$ for the Anderson model in Eq.~(\ref{deltaHLR}), Ref.~\onlinecite{logan} appeared which found the same result. This work was supported by the Government of British Columbia (JM), the A.V.~Humboldt Foundation (ES), the Natural Sciences and Engineering Research Council of Canada (IA), and the Canadian Institute for Advanced Research (JM, ES, IA).


\begin{thebibliography}{99}
\bibitem{footnote01} Only the lowest $M$ eigenvalues are kept at each iteration to prevent the Hilbert space from growing exponentially with increasing $N$.  We take $M = 3500$ (only counting those states that are not related to others by symmetry so that we are really considering more than 3500 states as described in ref.~\onlinecite{Jones}) and find that our results do not change if more states are kept.
\bibitem{footnote04} We obtain $K_c$ numerically to 8--9 significant digits.
\bibitem{footnote03} We use the extrapolated points rather than all of the data to avoid discrepancies due to deviations of $T^*$ away from $(\nu V_{LR})^2$ behaviour for large $\nu V_{LR}$.
\bibitem{erratum} Ref.~\onlinecite{Sela09Noise} contains a typo in the formula for $s[\alpha]$ below eq. 8; $\alpha$ should be removed from its RHS.
\end{thebibliography}
\end{document}